\documentclass[preprint,12pt]{elsarticle}

\usepackage[final]{changes}
\usepackage{amsmath} 
\usepackage{amssymb}
\usepackage{amsfonts}
\usepackage[export]{adjustbox}
\usepackage{graphicx}
\usepackage{url}
\usepackage{subcaption}
\usepackage{tabularx}
\usepackage{xcolor}
\usepackage{listings}
\usepackage[normalem]{ulem}
\usepackage{csquotes}
\usepackage{hyperref}
\usepackage{doi}
\usepackage{multirow}
\usepackage{booktabs}  
\usepackage{tikz}
\usepackage{tikz-uml}
\usetikzlibrary{arrows.meta}
\usepackage{pgfplots}
\tikzset{umlclass style/.style={
    draw, fill=yellow!20, 
    minimum width=3.25cm, minimum height=1cm, 
    font=\small\bfseries, align=center
}}

\journal{Nuclear Physics B}

\begin{document}

\begin{frontmatter}



\title{Employing Continuous Integration inspired workflows for benchmarking of scientific software - a use case on numerical cut element quadrature}

\author[label1]{Teoman Toprak\corref{mycorrespondingauthor}}
\ead{toprak@fdy.tu-darmstadt.de}
\author[label2]{Michael Loibl\corref{mycorrespondingauthor}}
\ead{michael.loibl@unibw.de}
\author[label3]{Guilherme H. Teixeira\corref{mycorrespondingauthor}}
\ead{teixeira@tugraz.at}
\author[label1]{Irina Shishkina}
\author[label1]{Chen Miao}
\author[label2]{Josef Kiendl}
\author[label3]{Benjamin Marussig}
\author[label1]{Florian Kummer}
 \affiliation[label1]{organization={Technical University of Darmstadt},
             addressline={Otto-Berndt-Str. 2},
             city={Darmstadt},
             postcode={64287},
             country={Germany}}

 \affiliation[label2]{organization={University of the Bundeswehr Munich},
             addressline={Werner-Heisenberg-Weg 39},
             city={Neubiberg},
             postcode={85577},
             country={Germany}}

 \affiliation[label3]{organization={Technical University of Graz},
             addressline={Technikerstraße 4/II},
             city={Graz},
             postcode={8010},
             country={Austria}}

\cortext[mycorrespondingauthor]{Corresponding author (these authors contributed equally to this work)}

\begin{abstract}
In the field of scientific computing, one often finds several alternative
software packages (with open or closed source code) for solving a specific
problem. These packages sometimes even use alternative methodological
approaches, e.g., different numerical discretizations. If one decides to
use one of these packages, it is often not clear which one is the best
choice. To make an informed decision, it is necessary to measure the
performance of the alternative software packages for a suitable set of test
problems, i.e. to set up a benchmark. However, setting up benchmarks
ad-hoc can become overwhelming as the parameter space expands rapidly.
Very often, the design of the benchmark is also not fully set at the start
of some project. For instance, adding new libraries, adapting metrics,
or introducing new benchmark cases during the project can significantly
increase complexity and necessitate laborious re-evaluation of previous
results. This paper presents a proven approach that utilizes established
Continuous Integration tools and practices to achieve high automation of
benchmark execution and reporting.
Our use case is the numerical integration (quadrature) on arbitrary
domains, which are bounded by implicitly or parametrically defined curves
or surfaces in 2D or 3D.
\end{abstract}




\end{frontmatter}



\section{Introduction}
\subsection{Benchmarks and Continuous Integration}
In the scope of this paper,
we define the term \emph{benchmarking}
as the performance evaluation and comparison of different algorithms or implementations for a specific problem.
This evaluation and comparison \replaced[id=TT]{typically involve}{is typically for} a set of test cases, i.e., \emph{benchmark problems}.

\replaced[id=TT, comment={since it is the first paragraph, I think it would be better to use a better language}]{In research software development, a common observation is}{It seems to be a general experience} that \added[id=TT]{even} small benchmarking projects quickly escalate in terms of complexity and effort.
\replaced[id=TT]{Furthermore, it is often observed}{It also seems to be the case} that benchmark projects \replaced[id=TT]{commonly}{always} require several iteration steps:
Even if \replaced[id=TT]{significant thought has been invested}{a lot of thought has been put} into the design at the outset, 
new aspects and questions usually arise after the planned tests have been completed, 
which then need to be addressed. 
The corresponding adaptations typically require the re-execution of many tests.
Since the parameter space of a cooperative benchmark is
-- \replaced[id=TT]{in a simplified sense}{somewhat simplified} --
the Cartesian product of the set of test candidates (in our case, the set of numerical libraries \replaced[id=TT]{selected for comparison}{we choose to compare})
times the set of test cases (in our case, the set of numerical integration problems we choose to solve),
the workload grows in quadratic fashion. Therefore, the manual execution of
benchmarks can become a very cumbersome process. This is especially true
since most of the time, after the first execution of the entire benchmarks, one
typically finds that the test cases might be changed or new test cases are added,
which then require the re-execution of all tests. To overcome this, a high degree
of automation is required, which allows the development of the benchmarking
process in an \emph{agile}, iterative manner.

In software engineering, the issue of executing many tests multiple times is
well-known, e.g., during the integration of new features or components into a
software package, and has been addressed by the practice of Continuous Integration
(CI). In this approach, developers regularly send their code changes to
a central repository, after which automated builds and tests are executed. Only
if all these tests are successful, changes are merged into the main code-base.
As a result, issues such as component incompatibility and the accumulation of
numerous untested changes can be mitigated by allowing developers to detect
them at an early stage. Regarding tests, one often makes a distinction between
\emph{unit tests} and \emph{regression tests}. Unit tests are small tests that check a single unit of code, while regression tests check whether the code reproduces the same
result as before after a change. The aim of both kinds of tests is to ensure that
the software is always in a working state.

In this work, we aim to provide guidelines and best practices for benchmarking
of computational scientific software, employing CI tools and practices.
Although our approach was developed for a specific use case, the underlying
principles are general and applicable across different projects.

In this work, we propose to use a tool developed for CI 
-- in particular, git pipelines --
to automate \deleted[id=TT]{to} the benchmarking of scientific software.
Pipelines allow the execution of arbitrary jobs on dedicated servers.
In a classical CI setup, these jobs would be the individual tests,
and the result would \added[id=ML]{be} a binary metric, i.e., either pass or fail.
This functionality, however, can be used to run benchmarks that produce, e.g., real-numbered results,
such as runtimes or error norms.

Most of the time, researchers would create such benchmark studies not for self-purpose,
but because they intend to build their own work on top of the benchmarked algorithms.
Therefore, the correctness of these algorithms is crucial.
\added[id=TT]{CI supports this demand by helping to maintain the software in a working and reliable state.}
\added[id=TT]{Moreover, CI-driven benchmarking significantly enhances the comparison and analysis of new developments against established references. 
With the help of automated and continuous testing, 
researchers can systematically examine key properties like accuracy, robustness, runtime, and scalability, enabling comprehensive evaluations. 
By reducing overhead and ensuring consistency, it also helps to improve research software quality, 
particularly benefiting smaller research groups and 
ultimately increasing its lifespan and usability.
}

\added[id=TT]{
\replaced[id=ML]{Externally developed benchmarks, i.e., those not set up by the main developer of the algorithm to be tested,}{This approach also} prevent\deleted[id=ML]{s} unintended biases in testing
\replaced[id=ML]{ since they}{, as externally developed benchmarks} restrict researchers' ability to subconsciously tailor tests toward their code's strength, 
fostering a fairer and more objective assessment.
Moreover, real-world experience frequently reveals that external users, unfamiliar with implementation details, may mistakenly perceive software as non-functional, despite it working correctly.
These scenarios highlight the importance of decoupling the code development from its verification\replaced[id=ML]{. This insight can be formulated as a key design principle:}{, a key design principle.}
By ensuring that testing procedures are independent of the code’s development, 
researchers can establish impartial verification processes that better reflect real\deleted[id=ML]{-world} usage scenarios.
}

\added[id=TT]{
Furthermore, CI-driven benchmarking promotes better scientific practices, aligning with the FAIR (Findable, Accessible, Interoperable, and Reusa- \allowbreak ble) principles \cite{Wilkinson2016}. 
Automated and regular testing ensures continuous monitoring of scientific software or research data, 
allowing researchers to systematically track changes, detect regressions, and validate results over time.
This does not only improve reproducibility but also facilitates transparency and reliability in research.
Moreover, it establishes a digital memory of the research process, 
preserving its evolution and ensuring backward compatibility in a rapidly changing software environment where libraries become obsolete, tools are replaced, and projects continuously evolve. 
This is also particularly beneficial in collaborative and dynamic research groups, 
where researchers frequently transition between tasks or teams, facilitating long-term accessibility of scientific workflows.
}
\subsection{Literature review}
\label{sec:Literature}

CI is a well-established practice in software development.
A meta-study, 
which reviews the impact of CI on software development,
is given by Soares et al.~\cite{soares_effects_2022}.
They conclude that CI positively impacts software development,
\added[id=ML]{whereby} the main effects are risk reduction and greater confidence in software (`quality assurance').
Furthermore, CI leads to the immediate fixing of broken code (`issues \& defects') 
and improvement in team communication.
A minor drawback can be that 
additional technical challenges are imposed on software development teams.

To our knowledge, 
there are only a few works that specifically discuss the use of CI tools 
in the domain of  scientific software development.
\deleted[id=TT]{We assume that most ?\\}
One of the first reports on the application of regression testing for numerical software
is given by Dubey et al.~\cite{dubey_ongoing_2015}, 
where they discuss their test environment for the verification of a specific multi-physics code.
In contrast to the work presented here,
they only consider a homogeneous Python environment; furthermore, their code is not open-source but only available upon request.
According to them, the code verification relies on three components:
First, a test suite that carefully balances between the widest possible code coverage (i.e., the relative amount of code covered by the test suite) and short runtimes.
Second, software to run the test suite.
Third, a tool to visualize and analyze the results.
In a follow-up work, Dubey \cite{dubey_good_2022}
gives a broader perspective and discusses generally good practices for scientific computing.
She also points out, just as we do, 
that most literature dedicated to good software engineering practices focuses on enterprise software
and that there is, unfortunately, only little literature dedicated to scientific computing. 
She further emphasizes the importance of automatic testing,
since following good practices generally improves the quality and credibility of scientific software
and the respective results achieved by its users. 
 Based on our experience, it is important to note that, compared to general software, scientific software has fundamental differences. For instance, testing is even more critical in scientific computing to ensure the correctness of numerical algorithms in comparison to, e.g., a graphical user interface. If there are hidden bugs that go unnoticed, this potentially compromises the validity of the findings. Additionally, the development of scientific software comes with challenges such as smaller development teams and high developer turnover.
A related presentation, partly by Dubey and her co-authors, is available online \cite{Bernholdt2022}.
They argue
that for research codes maintained over a long period of time,
the ``investment'' in sophisticated software engineering practices 
such as testing requires more work in the beginning but pays off over time.
In their presentation, they especially address the issues of 
designing software for flexibility and extensibility,
collaborative software development,
and the design of testing strategies for complex software systems.
They again emphasize the importance of
testing, since it guarantees the reproducibility of published results.

Maric et al.~\cite{maric_pragmatic_2023}
propose\deleted[id=ML]{s} a workflow model based on CI for research software engineering to improve software sustainability and reproducibility.
They promote version control, testing, and the use of \emph{merge requests}:
There, the main (or master) branch is protected, i.e., users cannot push directly to this branch.
Instead, users have to create a so-called merge request in order to merge their own contributions from another Git branch into the main branch.
The merge is only performed if all tests in the CI pipeline are successful.
With respect to research software development, they 
especially \replaced[id=TT]{demonstrate}{show} ways of cross-linking software with data and publications 
in ways that comply with data management obligations required by funding agencies, e.g., how digital object identifiers can be assigned to research software.

So far, the most comprehensive meta-study on testing practices for research software is given by Eisty et al.~\cite{eisty_testing_2025}.
There, 
a survey has been carried out among developers dealing with the general issue of  ``testing in research software.''
One of their conclusions, w.r.t. ``Community Building and Knowledge Sharing,''
is that:
``The study underscores the importance
of knowledge sharing and community building within the research software development
community. Initiatives aimed at sharing best practices, lessons learned, and tools for
testing research software can foster collaboration and accelerate improvements in testing practices'' \cite[pg. 36]{eisty_testing_2025}.
The work we present here can be seen as a contribution to this goal.
\replaced[id=TT]{Furthermore, they list the following points}{They further conclude} \cite[pg. 38]{eisty_testing_2025}:
\comment[id=ML]{The complete bullet point list is a direct citation. Should we place quotation marks somewhere?}
\comment[id=FK]{Added the block-quote environment; I think this is the correct way to do it?}
\begin{displayquote}
    \begin{itemize}
    \item Testing Practices Lag Behind Industry: Research software testing is less systematic
    and less automated than in industry, with significant gaps in knowledge and tool adoption.
    \item Unique Challenges: The \emph{oracle problem}, i.e., to decide whether the output of the test is correct or not,
    and test case design are major challenges,
    especially in complex, non-deterministic systems.
    \item Resource Limitations: Smaller teams and limited budgets lead to ad-hoc testing, compromising
    software quality and reliability.
    \item Need for Specialized Tools: Current tools do not fully address the unique needs of
    research software, highlighting a demand for more tailored solutions.
    \item Cultural Shift Needed: Testing is often an afterthought in research software. A change
    in mindset is required, integrating testing as a core part of development.
    \item Educational Opportunities: Integrating software testing education into research curricula
    can help close the knowledge gap and improve future practices.
    \item Community Building: Strengthening collaboration and knowledge sharing in the
    research software community can help address tool fragmentation and improve sustainability.
    \end{itemize}
\end{displayquote}

We are certain  that our work contributes to the goals discussed above:
The testing practices we propose rely on established industry tools, aiming to reduce the lag behind industry.
The use of established industry tool-chains
generally helps especially smaller teams to overcome configuration overhead and other limitations,
since the respective tools are very mature and easy to set up.
For instance, 
when GitLab is used,
this can even help to overcome resource limitations.
For example, some team can easily build up a cluster of significant size for testing,
by configuring  
 all available workstations in a group as runners (i.e., cluster nodes),
with only a few minutes of configuration effort per machine.
An agile approach to testing, as we propose, can also help to propagate a cultural shift:
Testing is easy to set up initially and can grow with the project.

This paper is organized as follows:
Section~\ref{sec:usecase} describes \emph{our specific use case}.
In Section~\ref{sec:general-framework}, we draw a broader picture
by sketching a \emph{general framework} from this use case, which the reader might apply to his application.

\section{Description of the use case}
\label{sec:usecase}

The main purpose of our use case is the evaluation of different approaches to numerical integration, 
also known as quadrature, \replaced[id=IS]{on}{of} cut elements.
Considering that multiple benchmarks are defined in a unified environment that accesses various codes utilizing different integration methods.
In this section, only a brief overview from a high level is given.
For details, the reader is referred to the respective source code repository available on GitHub
\cite{CutElementIntegration2025}
under an open-source BSD license.

\paragraph{Background and motivation - Galerkin methods on cut element meshes}

\added[id=ML]{The motivation behind our use case is \added[id=TT]{numerical methods for engineering problems}, e.g., 
Finite Element Method (FEM), Isogeometric Analysis (IGA), or Discontinuous Galerkin Methods (DGM). 
The common denominator of FEM, IGA, and DGM is that they are so-called Galerkin methods, 
which, at their core, 
require numerical integration in finite domains, so-called elements (or cells). The accumulation of elements yields a mesh representing the analysis domain. Generating meshes can become challenging for complex geometries.} 

\begin{figure}
    \centering
	\begin{subfigure}{0.3\textwidth}
		\includegraphics[width=\textwidth]{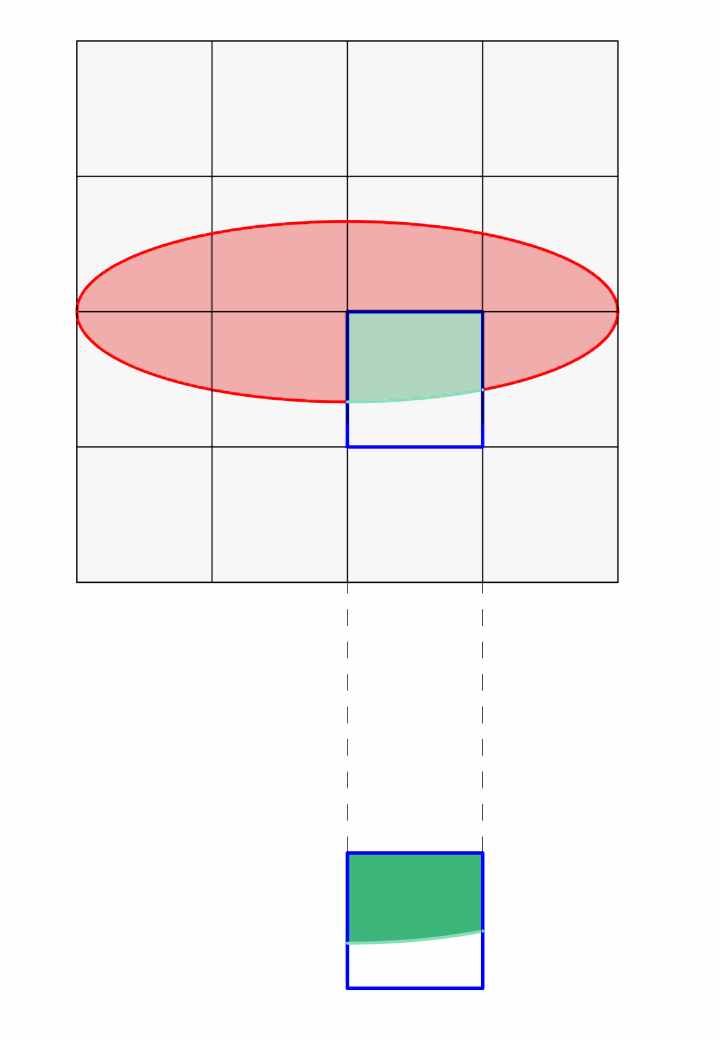}
	\end{subfigure}
	\hfill
	\begin{subfigure}{0.55\textwidth}
		\includegraphics[width=\textwidth]{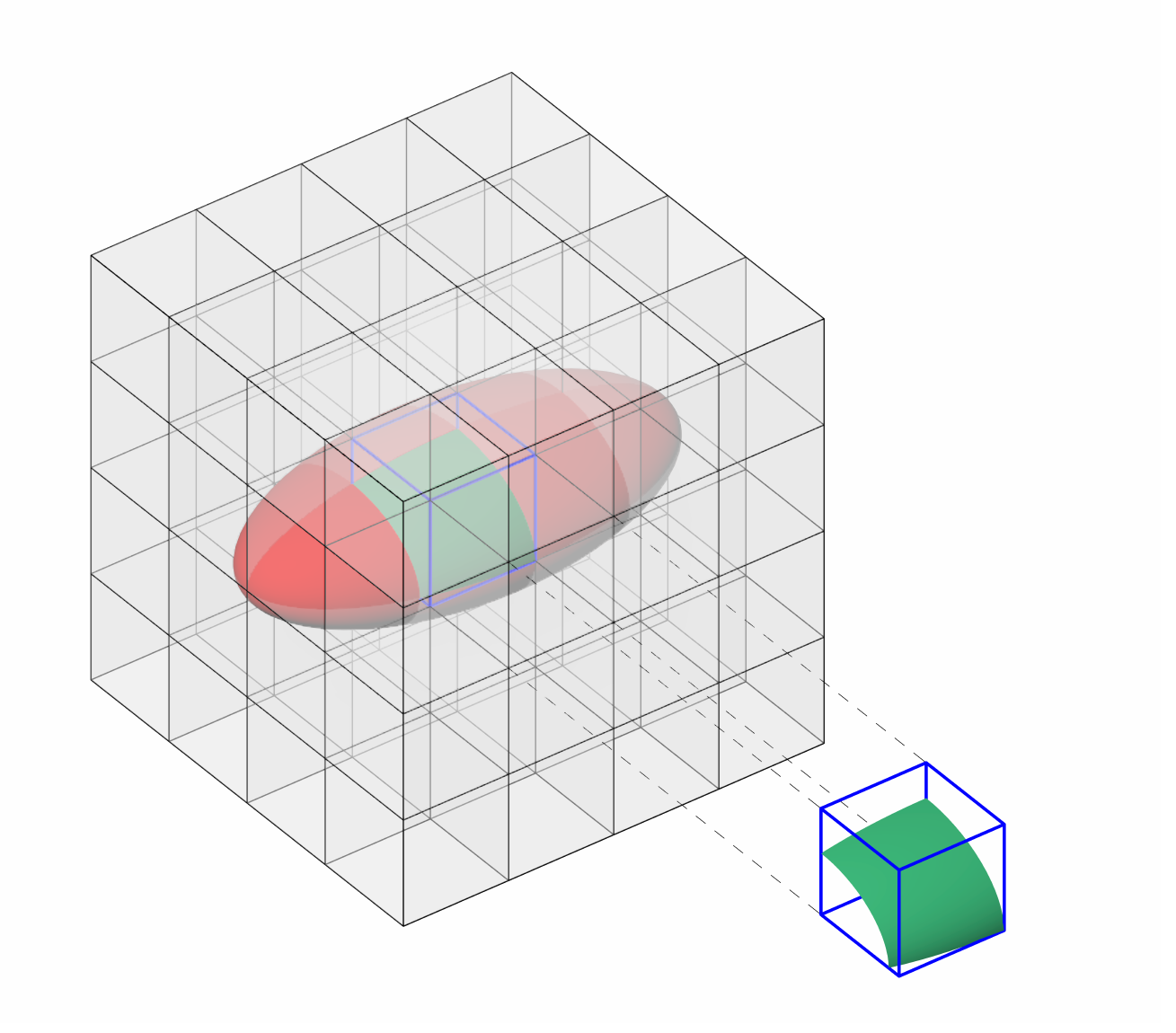}
	\end{subfigure}
    \caption{Illustration of cut elements, which are created by intersecting
    a -- \replaced[id=IS]{typically}{very often}, Cartesian -- background mesh with an embedded interface.
    Left:  2D case. Right: 3D case.}
    \label{fig:cut-cells}
\end{figure}

\added[id=ML]{Geometrically unfitted methods simplify the potentially complex meshing process in such Galerkin methods. In such approaches, a domain boundary or interface is embedded in a -- typically Cartesian -- background mesh, as shown in Figure \ref{fig:cut-cells}. In several applications, allowing elements to be cut by an interface can be beneficial with respect to further reasons.
For example, this enables tracking  moving domain boundaries or cracks 
without a continuous reconstruction of the elements. However, the numerical integration must be able to deal with cut and uncut elements as a consequence of the embedding.}

\added[id=ML]{By applying an integral transformation, the elements not intersected by the interface can be mapped to a reference element, e.g., $(-1,1)^2$ or $(-1,1)^3$ in the case of quadrilaterals or hexahedrons, respectively. A predefined quadrature rule (e.g., Gauss quadrature) can then be used for the reference element, making integration \replaced[id=IS]{relatively straightforward}{rather easy}}.

\added[id=ML]{Performing an \emph{accurate and efficient} integration on cut elements is much more challenging since such a general mapping of the active part of a cut element to a reference element is not known a priori.} \added[id=ML]{Various authors \cite{Mueller2013,Zander2014,Saye2015,Kummer2017,Gunderman2021,Messmer2022,Saye2022} have \replaced[id=IS]{proposed}{come up with} different approaches 
for creating cut element quadrature rules \replaced[id=IS]{dynamically}{on the fly} during the simulation
and 
respective software libraries are available, even under open-source licenses.}

\deleted[id=BM]{
I suggest to reduce the elimination of meshing focuse of the previous paragraph a bit. Consider:\\
For these simple cells, numerical integration is rather easy. 
One can apply an integral transformable to the reference element, using the aforementioned $\mathcal{C}^1$ mappings.
Then, a predefined quadrature rule -- e.g., Gauss quadrature -- can be used for the reference element.
However, in several applications, allowing cells to be cut by an interface, as shown in Figure \ref{fig:cut-cells}, can be beneficial. For example, this enables tracking moving domain boundaries or cracks without a continuous reconstruction of the cells.
At the same time, it is much more challenging to perform \emph{accurate and efficient} numerical integration in cut elements
since a general mapping of the active, cut part of the cell to a reference element is not known.
}
\comment[id=FK]{ok, taken Benjamins suggestion}
\deleted[id=TT]{I would focus on geometrically unfitted methods, rather than Galerkin methods as one can also do cut celement quadrature for FVM or hybrid schemes.}


\paragraph{The task at hand}
In mathematical terms, 
we want to
compute boundary and interior integrals
 of a function $f$ over a domain $\Omega$:
\[
    \oint_{\Omega \cap I } f \, \text{d}S, \quad \int_{\Omega \cap A} f \, \text{d}V .
\]
Here, $I$ is the `$D-1$'-dimensional, sufficiently smooth boundary,
also called the \emph{interface}, 
of the $D$-dimensional \emph{enclosed domain} $A \subset \Omega \subset \mathbb{R}^D$. We consider two options to describe the interface:
\begin{itemize}
    \item \added[id=ML]{Implicit representation: $\phi(\mathbf{x})$. $\phi$ defines a level-set function, which is zero on the interface.
	In this case, $A$ is the domain in which $\phi(\mathbf{x}) \leq 0$.}
	
	\item \added[id=ML]{Parametric representation: $s \mapsto \mathbf{x}(s)$. 
	In this work, we assume that the parametric interface representations are given as closed oriented loop of NURBS curves.}


\end{itemize}

In the benchmarks conducted for this paper, we employ the open-source software libraries for cut element quadrature listed in Table ~\ref{tab:quadrature-libraries}. It should be noted that all of these libraries support 
only one of the aforementioned interface representations,
either implicit or \replaced[id= TT]{parametric}{explicit} interface representation. 
\comment[id=BM]{Should the interface type be shown in the table above?
}
\comment[id=FK]{I added it. I hope its not to cramped. I really would like to keep the table within the text, not as a separate float, for readbility.}

\begin{table}[htbp]
  \caption{Cut-cell quadrature libraries and their characteristics}
  \label{tab:quadrature-libraries}
  \centering
  \begin{tabular}{|
      >{\raggedright}p{1.6cm} |
      p{1.7cm} |
      p{2.2cm} |
      p{1.8cm} |
      p{1.5cm} |
      p{1.8cm} |
      p{1.2cm} |
    }
    \hline
    \textbf{Name}
      & \textbf{Interface repr.}
      & \multicolumn{1}{>{\raggedright\arraybackslash}p{2.5cm}|}{\textbf{Application}}
      & \textbf{Lang.}
      & \textbf{OS}
      & \textbf{License}
      & \textbf{Ref.} \\
    \hline

    Algoim
      & Implicit
      & \multicolumn{1}{>{\raggedright\arraybackslash}p{2.5cm}|}{High-order multi-scale multi-physics modeling}
      & C++
      & Linux
      & BSD-3
      & \cite{Saye2015,Saye2022}, \cite{Algoim} \\
    \hline

    BoSSS
      & Implicit
      & \multicolumn{1}{>{\raggedright\arraybackslash}p{2.5cm}|}{High-order DG for multi-physics flows}
      & C\#
      & Windows
      & Apache 2.0
      & \cite{Mueller2013,Kummer2017}, \cite{BoSSSRepo} \\
    \hline

    FCMLab
      & Implicit
      & \multicolumn{1}{>{\raggedright\arraybackslash}p{2.5cm}|}{p-FEM, FCM}
      & MATLAB
      & Cross-\newline platform
      & GPL-3.0
      & \cite{Zander2014}, \cite{FCMLabRepo} \\
    \hline

    Ginkgo
      & Param.
      & \multicolumn{1}{>{\raggedright\arraybackslash}p{2.5cm}|}{Manycore sparse linear systems}
      & C++
      & Linux
      & BSD-3
      & \cite{antolin_isogeometric_2019}, \cite{GinkgoRepo} \\
    \hline

    Nutils
      & Implicit
      & \multicolumn{1}{>{\raggedright\arraybackslash}p{2.5cm}|}{FEM, IGA,\newline FCM, multi-physics}
      & Python
      & Cross-\newline platform
      & MIT
      & \cite{divi_error-estimate-based_2020,divi2022residual}, \cite{NutilsRepo} \\
    \hline

    ngsxfem
      & Implicit
      & \multicolumn{1}{>{\raggedright\arraybackslash}p{2.5cm}|}{Active-mesh XFEM with level-set quadrature, space–time FEM}
      & C++/\newline Python
      & Cross-\newline platform
      & LGPL-3.0
      & \cite{Lehrenfeld2016,lehrenfeld_higher_2017}, \cite{ngsxfemRepo} \\
    \hline

    QuaHOG
      & Param.
      & \multicolumn{1}{>{\raggedright\arraybackslash}p{2.5cm}|}{Higher-order quadrature  with arbitrary integrands}
      & MATLAB
      & Cross-\newline platform
      & BSD-3
      & \cite{Gunderman2021}, \cite{QuaHOGRepo} \\
    \hline

    QuaHOG- \allowbreak PE
      & Param.
      & \multicolumn{1}{>{\raggedright\arraybackslash}p{2.5cm}|}{Higher-order quadrature, exact for any polynomial integrand}
      & MATLAB
      & Cross-\newline platform
      & BSD-3
      & \cite{Gunderman2021}, \cite{QuaHOGRepo} \\
    \hline
  \end{tabular}
\end{table}

\paragraph{A unified driver interface for all libraries to investigate}
In a first step,
a Git repository was set up (in the following referred to as the `\emph{main repository}'),
where each of these third-party libraries was included as a submodule.

\begin{figure}
    \centering
	\scalebox{0.8}{ 

\begin{tikzpicture}
 \umlclass[x=0, y=0.6]{AbstractIntegrator}{
        + Name : string \\
        + InterfaceType : string (``parametric'' | ``implicit'') \\
        + OperatingSystem : string (``Linux'' | ``Windows'') \\
        + SupportedDimensions : string (``2D'' | ``3D'') \\
        ... \\
    }{
        + computeArea2D(BenchmarkTest)  \\
        + computeVolume3D(BenchmarkTest)  \\
        ...
    }

    \umlclass[x=0.5, y=-6]{Testcase}{
        - dim : int \\
        + domain \\
        + interface \\
        + references 
    }{
        + Testcase(domain, interface, references, ...)
    }

    


\node[umlclass style] (Quaddata) at (7, -0.7) {QuadratureData};

\draw [decorate, decoration={brace, amplitude=10pt}] 
    ([yshift=-0.9cm,xshift=-1.2cm] AbstractIntegrator.east) -- ([yshift=-1.7cm,,xshift=-1.2cm] AbstractIntegrator.east) ;
\draw[-{Triangle[scale=1.5]}]
    ([yshift=-1.3cm,xshift=-0.9cm] AbstractIntegrator.east)  -- (Quaddata.west);

\draw [decorate, decoration={brace, amplitude=10pt}] 
    ([yshift=0.5cm,xshift=1.5cm] AbstractIntegrator.south) -- ([yshift=0.5cm,,xshift=-0.6cm] AbstractIntegrator.south) ;
\draw[-{Triangle[scale=1.5]}]
    ([yshift=0.2cm,xshift=0.45cm] AbstractIntegrator.south) -- ++(0,-0.5) -- ++(0,-1.5) -- ([xshift=-0.05cm] Testcase.north);

\node[umlclass style] (Domain) at (7, -5.5) {Domain};
\node[umlclass style] (Interface) at (7, -7) {Interface};
    
\node[umlclass style] (AlgoimIntegrator) at (0,-9) {AlgoimIntegrator};
\node[umlclass style] (BoSSSIntegrator) at (0,-10.5) {BoSSSIntegrator};

\draw[-] (AlgoimIntegrator.west) -- ++(-2,0) ; 
\draw[-{Triangle[scale=1.5]}] (BoSSSIntegrator.west) -- ++(-2,0) -- ++(0,8) -- ([xshift=-3cm] AbstractIntegrator.south);

\draw[-] ([yshift=-1cm,,xshift=-2cm] BoSSSIntegrator.west) --([yshift=0cm,xshift=-2cm] BoSSSIntegrator.west) node[below=1cm, right] {$\boldsymbol{...}$};

\draw[-{Triangle[scale=1.5]}] ([xshift=-2.5cm, yshift=0.25cm] Testcase.east) -- (Domain.west) ;
\draw[-{Triangle[scale=1.5]}] ([xshift=-2.5cm, yshift=-0.25cm] Testcase.east) -- (Interface.west);

\end{tikzpicture}

}
    \caption{Class diagram of the MATLAB control code:
The unified driver interface for all libraries is defined in the \texttt{AbstractIntegrator} class,
from which all integrator classes are derived.
Various \texttt{compute*} methods actually execute the numerical integration.
The input to these methods is always a \texttt{Testcase} object, 
which contains the computational domain as well as the interface representation.
The \texttt{Testcase} objects always actually support both representations,
to guarantee maximal compatibility.}
    \label{fig:matlabUML}
\end{figure}
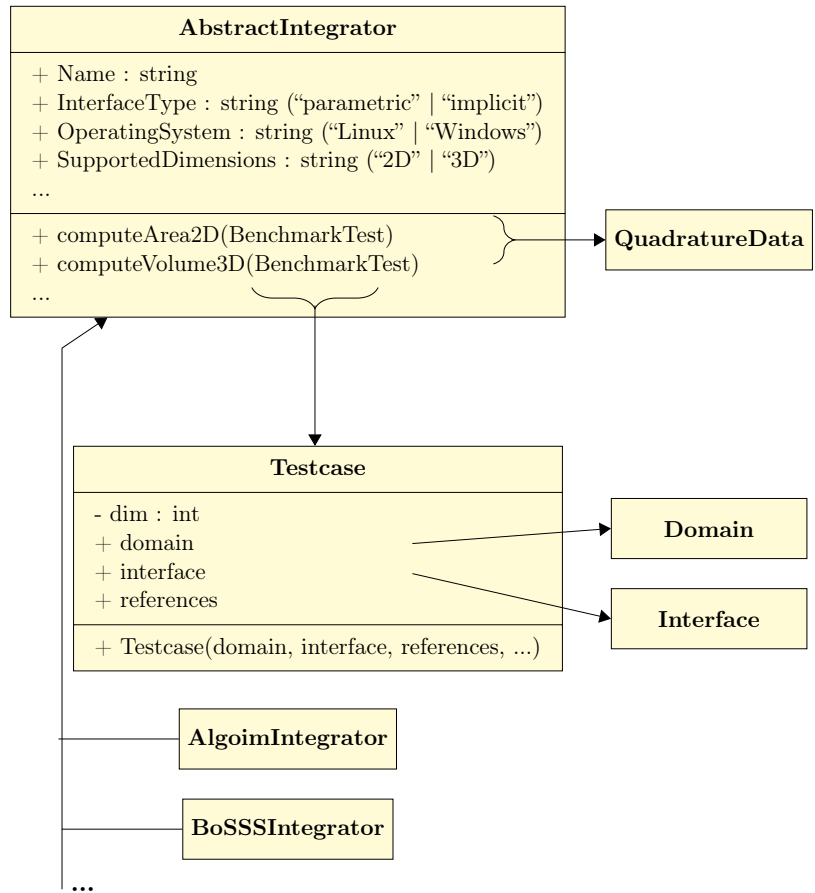
\comment[id=ML]{I like the idea of the graph in Figure \ref{fig:matlabUML}. I would add "dim" and "references" to "TestCase" (I found that what we call "references" is often called "baselines" in literature, e.g., in \cite{Bernholdt2022}). Furthermore, I would add "QuadratureData" as class showing unified output. Further outputs are plots of the quadrature points with geometry and background mesh, plots of the error and tables ".csv" with the errors and number of points. These tables are actually used then for the comparison with the references for the pipelines. If you think that this makes sense, one could add that here. $\Longrightarrow$ already partly addressed by Toprak $\Longrightarrow$ I would discuss to discuss this in more detail with you (Toprak) for the journal submission.}

Next, a control code to define and execute the benchmark tests was developed in MATLAB.
The class structure is shown in Figure \ref{fig:matlabUML}. 
The key idea is to define a common interface for all libraries.
\added[id=BM]{Figure \ref{fig:abstractclass} shows the abstract class for integrators of our use case. Note that the class does not provide any functionality but specifies the properties, functions, and outputs required for all derived integrator classes to be compatible with the test environment.}

\begin{figure}
	\centering
\lstset{
    language=Matlab,
    basicstyle=\ttfamily\scriptsize,
    keywordstyle=\bfseries\color{blue},
    commentstyle=\itshape\color{green!50!black},
    stringstyle=\color{red},
    breaklines=true,
    frame=single,
    showstringspaces=false
}

\begin{lstlisting}
classdef (Abstract) AbstractIntegrator 
   methods (Abstract)
       % methods to manage the different integrators
       out = Name;                % string specifying the integrator name
       out = InterfaceType;       % "parametric" or "implicit"
       out = OperatingSystem;     % "Linux" and/or "Windows"
       out = SupportedDimensions; % "2D" and/or "3D"
       out = IsAccessible;        % check if implementation is in 'codes' subfolder  
       out = PropertyString;      % string encoding the properties of the integrator
       % methods for integration
       result = computeArea2D(BenchmarkTest);
       result = computeVolume3D(BenchmarkTest);
       result = computeInterfaceCurveLength(BenchmarkTest);
       result = computeInterfaceSurfaceArea(BenchmarkTest);
       result = computeAreaViaFlux2D(BenchmarkTest);
       result = computeVolumeViaFlux3D(BenchmarkTest);
   end
end
\end{lstlisting}	
\caption{\added[id=BM]{MATLABs abstract class for the integrator or our use case.}}
\label{fig:abstractclass}
\end{figure}

Here, the devil is in the details:
In our case, most of the libraries only support one of the two interface representations,
either implicit or \replaced[id= TT]{parametric}{explicit}.
Issues like this could be resolved by a more refined class structure,
e.g., by introducing an \texttt{AbstractImplicitIntegrator} and an \texttt{AbstractParametric \\ Integrator} class.
However, for the sake of consistency, 
we decided to use a base class for all integrators
and a base class for all test cases.
The integrators expose a member \texttt{InterfaceType} that informs 
which of the two interface representations is supported.
The test cases, on the other hand, have to support both interface representations,
so that they can be applied for all integrators.
The bottom line is that in some software design decisions,
one has to balance strictly following a paradigm, 
e.g., using object-oriented programming 
but including exceptions such as if-else-statements.
While the right amount of exceptions can definitely aid simplicity and save code-lines,
too many exceptions can lead to a code that is hard to understand and maintain.


\paragraph{Continuous Integration via GitLab}
The MATLAB code serves the execution of the tests and the evaluation of the results. Several test suites are set up, each collecting a group of benchmark tests for a specific functionality to be tests, e.g., domain/boundary integrals in 2D/3D. Furthermore, each benchmark test has a 'full' and 'quick' option, where the latter executes only a subset of the test's configuration variations. This modularization lets developers get quick feedback if their code changes affect a specific benchmark test. Nevertheless, running all tests individually on a local machine last several hours and would hinder working on that machine. Further, some libraries require compilation or installation. Moreover, there is the problem of heterogeneity. Not all libraries can run on the same operating system, e.g., the C\#-based BoSSS library can only be linked to MATLAB on Windows (cf. Table \ref{tab:quadrature-libraries}). One could employ some form of virtualization here, but this has to be set up individually for every contributor to the code individually.

Above all, especially in a collaborative work,
it is necessary to ensure that the contributors do not break the code of the others;
this is a typical source of friction in collaborative projects.

\begin{figure}
    \centering
    \includegraphics[width=0.99\textwidth]{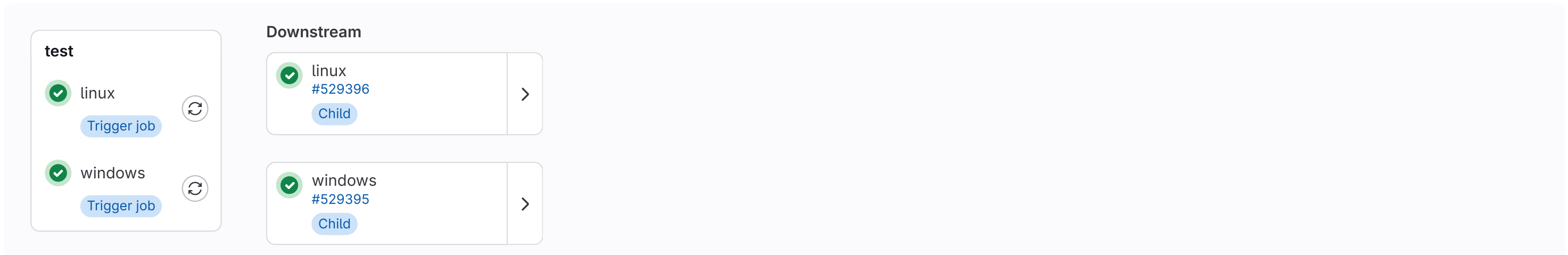}
    \includegraphics[width=0.99\textwidth]{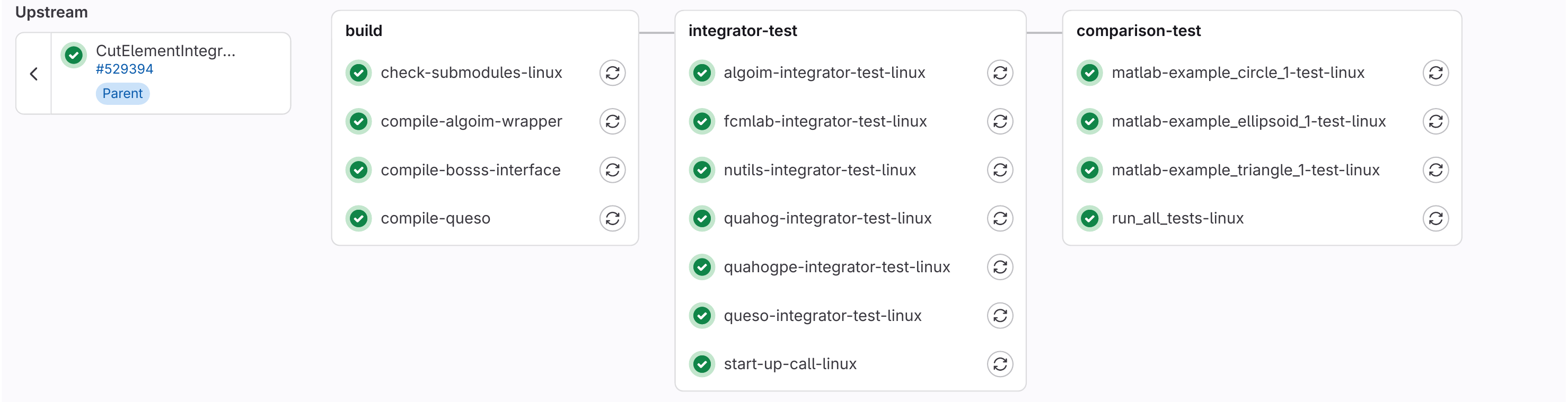}
    \includegraphics[width=0.99\textwidth]{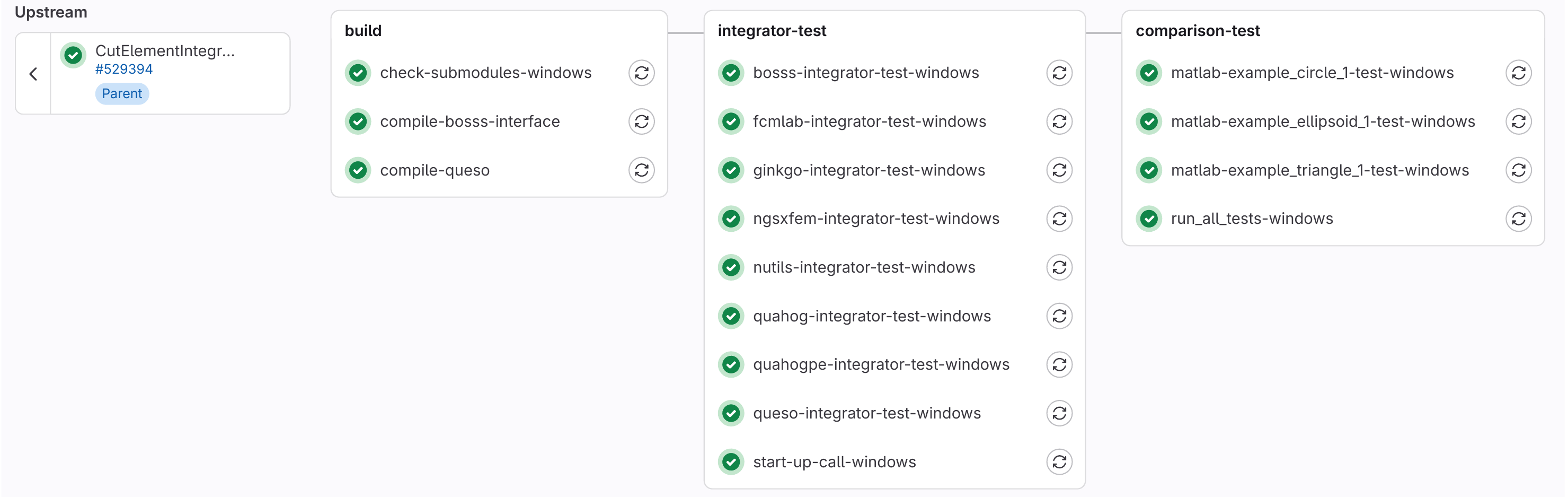}
    \caption{
        Visualization of a successful pipeline in GitLab.         
    }
    \label{fig:GitLab}
\end{figure}

To overcome these issues, we employed GitLab \emph{pipelines} (alternative platforms such as GitHub\added[id=TT]{ or Azure DevOps} could be used equivalently).
Therefore, the \texttt{main} branch (very often also called the \texttt{master} branch) is protected,
i.e., no one can push directly to this branch.
Instead, one has to create a so-called \emph{merge request} in order to merge  contributions 
from some Git branch into the \texttt{main} branch.
For the merge request to be accepted, the GitLab pipeline must be successful, i.e., none of its tests failed.
The pipeline is defined in a so-called YML-script 
in the repository (always named \texttt{.gitlab-ci.yml}) which specifies the individual jobs to be executed. 
Figure \ref{fig:GitLab} shows a visualization of a  GitLab pipeline running  two parallel paths (for Linux and Windows) through four  steps: upstream, build, integrator-test, and comparison-test. When all jobs complete successfully, green checkmarks appear, as shown here.

An essential feature of GitLab is that it allows us to execute jobs on our `own' hardware while still relying on the GitLab web-based service, which is, in our case,
hosted by Graz University of Technology.
This is achieved via so-called \emph{GitLab runners}:
A runner is a background program (also known as demon or service on Linux/Unix, service on Windows) that can be installed on any machine.
At installation time, the runner is registered with the GitLab server.
When a GitLab pipeline is executed, the GitLab server then sends the jobs to the runner, which executes them. Each runner can be tagged with a user-defined label, e.g., \texttt{linux-runner} or \texttt{windows-runner}.
In the YML-script, one  can assign jobs to runners by specifying the respective label.

\begin{figure}
    \centering
    \includegraphics[width=0.99\textwidth]{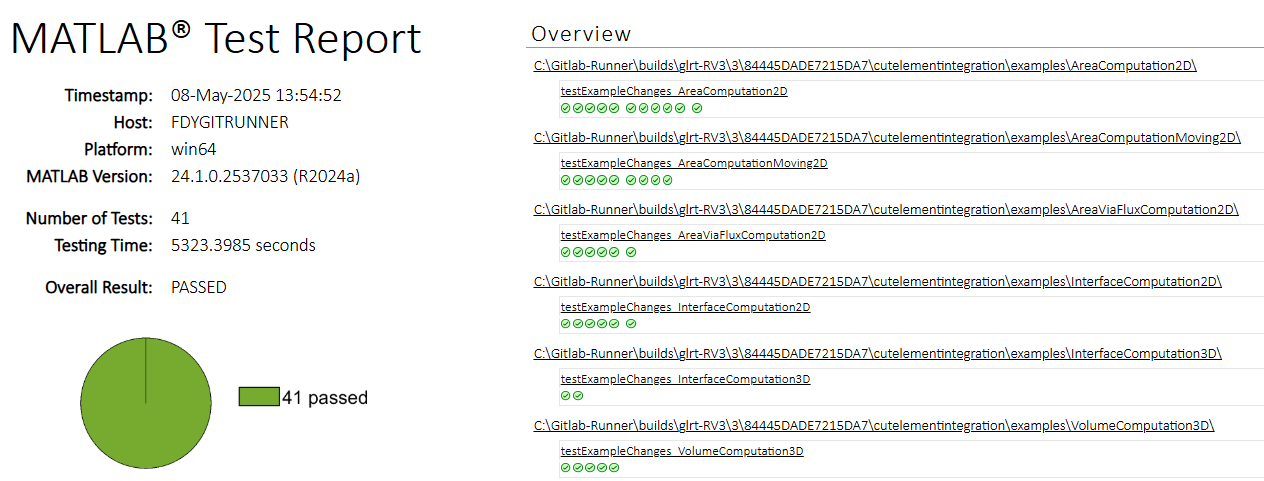}
    \caption{An example of a MATLAB test report.}
    \label{fig:Matlab_report}
\end{figure}

Finally, the MATLAB scripts produce benchmarking results, e.g., tables and plots. 
In the GitLab pipeline, we use the \texttt{artifacts} feature to store these results.
For each job, one can define a set of files that are stored as artifacts.
Then, two kinds of things can be done:
the artifacts can be downloaded from the GitLab web interface, e.g., to be used in a publication.
Or, the artifacts can be used in subsequent jobs, e.g., gather a large result table from the results of individual jobs. Additionally, these artifacts include automatically generated MATLAB test reports after each pipeline execution. Figure \ref{fig:Matlab_report} illustrates an example of such a report, demonstrating its structure and content.

\paragraph{Example results}
The test cases for benchmarking should cover a sufficiently broad variety of problems to guarantee code integrity through the project workflow. The following aspects should be varied in the context of our use case for proper testing: background meshes, boundary shape, integral type (i.e., boundary or domain integral), and integrands. Furthermore, the robustness of the integration tools is important and should be tested. This can be done for example by moving geometries through the background mesh to provoke unexpected cutting situations. In the following, two examples with a circular disk are presented: a convergence study under mesh refinement is performed in the first case, and a shift of the geometry through a fixed mesh is investigated in the second case. Further examples comparing the Ginkgo and the Algoim code can be found in \cite{Teixeira2024}.

\begin{figure}
	\centering
	\begin{subfigure}{0.32\textwidth}
		\includegraphics[width=\textwidth]{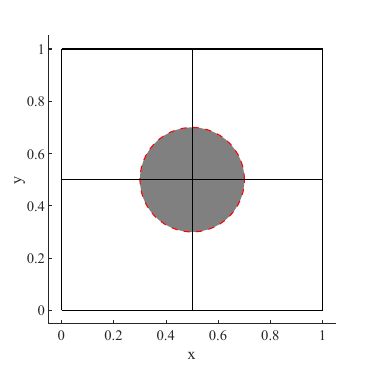}
		\caption{Problem setup \deleted[id=TT]{Enclosed domain in grey.}}
		\label{fig:1example_setup}
	\end{subfigure}
	\hfill
	\begin{subfigure}{0.32\textwidth}
		\includegraphics[width=\textwidth]{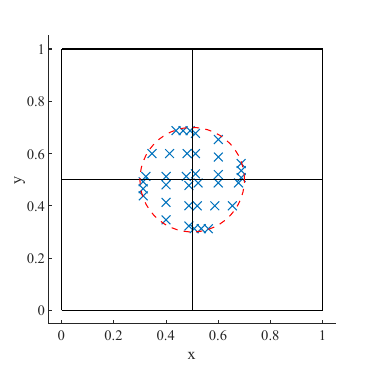}
		\caption{Algoim}
		\label{fig:1example_algoim}
	\end{subfigure}
	\hfill
	\begin{subfigure}{0.32\textwidth}
		\includegraphics[width=\textwidth]{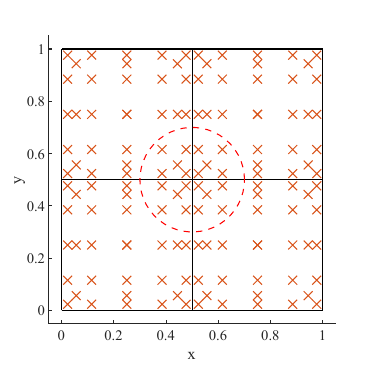}
		\caption{BoSSS}
		\label{fig:1example_bosss}
	\end{subfigure}
	\begin{subfigure}{0.32\textwidth}
		\includegraphics[width=\textwidth]{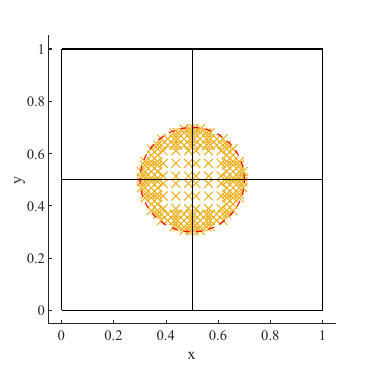}
		\caption{FCMLab}
		\label{fig:1example_fcmlab}
	\end{subfigure}
	\hfill
	\begin{subfigure}{0.32\textwidth}
		\includegraphics[width=\textwidth]{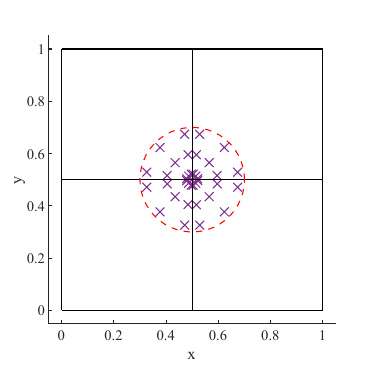}
		\caption{Ginkgo}
		\label{fig:1example_ginkgo}
	\end{subfigure}
	\hfill
	\begin{subfigure}{0.32\textwidth}
		\includegraphics[width=\textwidth]{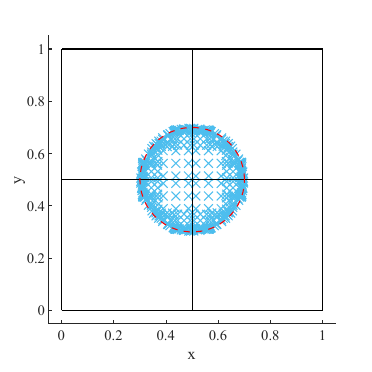}
		\caption{Nutils}
		\label{fig:1example_nutils}
	\end{subfigure}
	\hfill
	\begin{subfigure}{0.32\textwidth}
		\includegraphics[width=\textwidth]{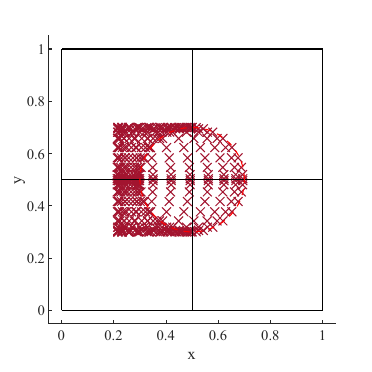}
		\caption{QuaHOG}
		\label{fig:1example_quahog}
	\end{subfigure}
	\begin{subfigure}{0.32\textwidth}
		\includegraphics[width=\textwidth]{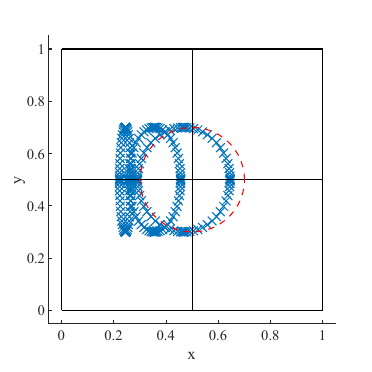}
		\caption{QuaHOGPE}
		\label{fig:1example_quahogpe}
	\end{subfigure}
	
	\caption{First test case. Circular disk with centre point $C=[0.5,0.5]$ and radius $R=0.2$. Quadrature points generated by different software libraries for a $2\times 2$ mesh are depicted as crosses. The interface is illustrated as red dashed curve.}
\end{figure}

\begin{figure} [!bth]
	\centering
	\includegraphics[width=1.5\textwidth, center]{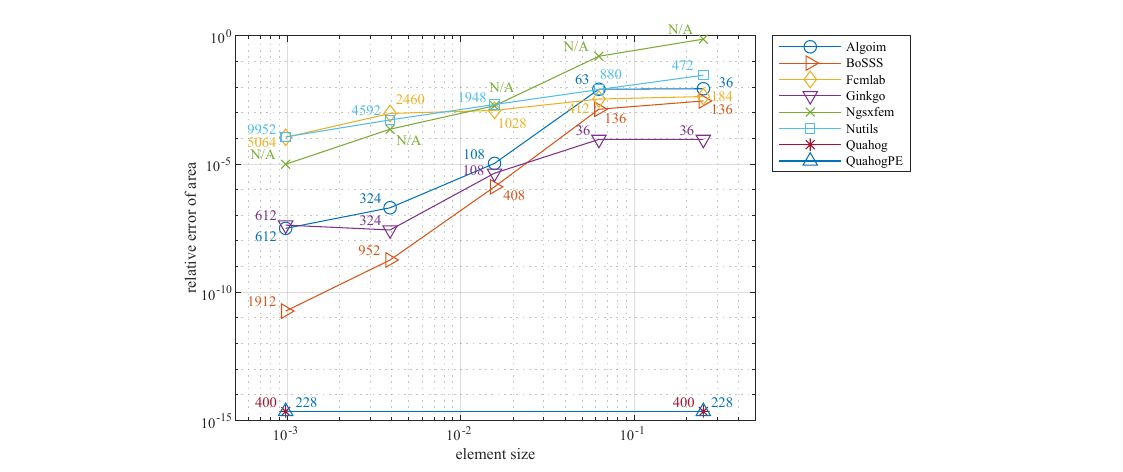}
	\caption{First test case. Convergence study of the relative error of the area under uniform mesh refinement. The number next to the markers show the number of used quadrature points.}
	\label{fig:1example_convergence}
\end{figure}

The setup of the first test case with an initial mesh of $2\times2$ elements is illustrated in Figure \ref{fig:1example_setup}. The resulting quadrature points for the different methods are illustrated in Figure \ref{fig:1example_algoim} to \ref{fig:1example_quahogpe}. The area of the circle is computed to check the geometrical accuracy of the proposed integration schemes. The mesh is uniformly refined up to $32\times 32$ elements. Figure \ref{fig:1example_convergence} illustrates the relative errors of the numerically determined area versus the exact reference solution for each integration scheme with respect to the element size, i.e., $1/n$, where $n$ is the number of elements. The graphs reflect the accuracy and performance of the related scheme and allow for assessing if an implementation complies with its underlying mathematical theory. The QuaHOG and QuaHOGPE codes apply a mesh-free approach and thus the resulting quadrature points are not directly comparable to the others. Therefore, we omitted the markers for the respective curve for some element sizes. 

\begin{figure} [!bth]
\centering
\includegraphics[width=1.5\textwidth, center]{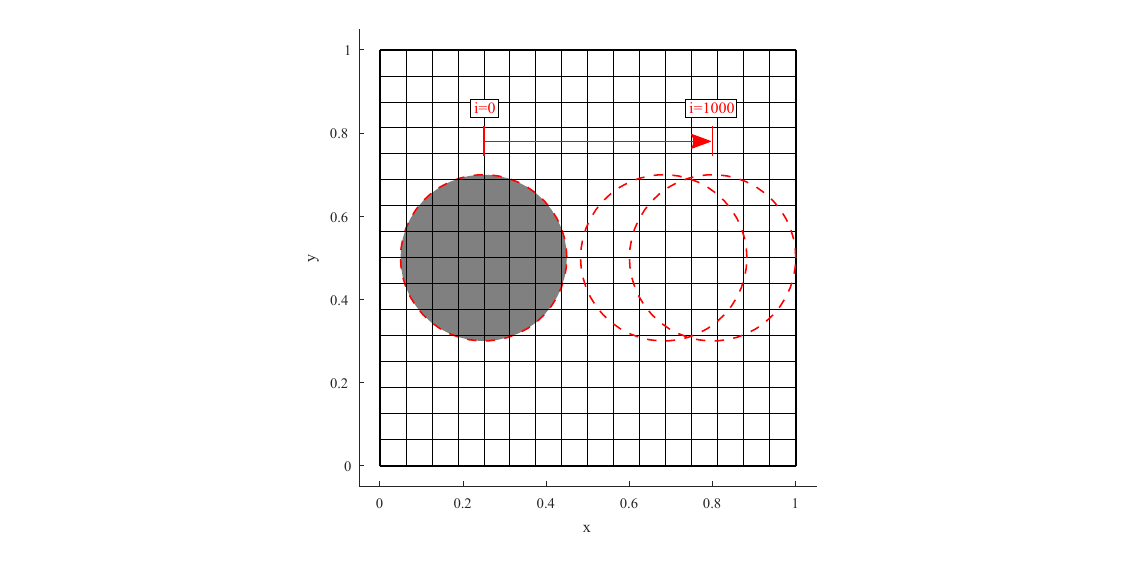}
\caption{Second test case. Shift of circular disk from left to right with 1000 steps. The center point initial position is $C=[0.25,0.5]$ and the radius is $R=0.2$. The interface is illustrated as red dashed curve.}
\label{fig:2example_setup}
\end{figure}

\begin{figure} [!bth]
	\centering
	\includegraphics[width=1.2\textwidth, center]{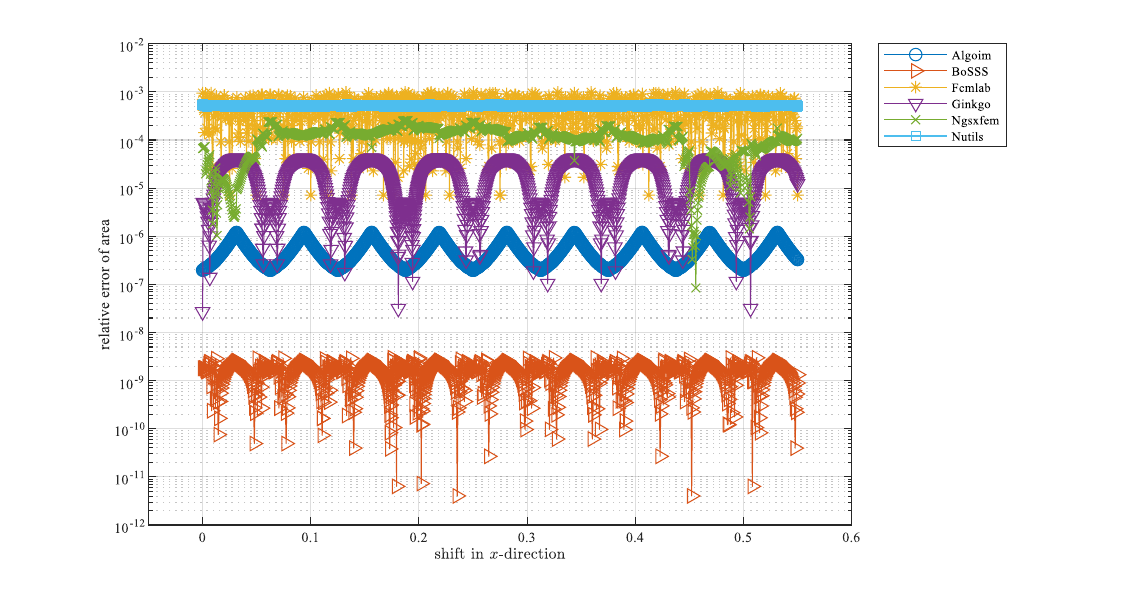}
	\caption{Second test case. Relative error for the problem depicted in Figure \ref{fig:2example_setup}.}
	\label{fig:2example_error}
\end{figure}

The setup of the second test case is illustrated in Figure \ref{fig:2example_setup}. Here, the circular disk is moved from left to right with 1000 steps in order to trigger various cutting situations. A fixed background mesh with $8 \times 8$ elements is considered. The area of the circle is again computed to check the geometrical accuracy of the proposed integration schemes. The relative errors are illustrated in Figure \ref{fig:2example_error}. Less oscillations indicate a better robustness of the code, whereby it has to be considered that the error plot is logarithmic. QuaHOG and QuaHOGPE are not considered in this example because their performance is not affected by a geometry shift as they are mesh-free methods.

The error plots and the plots of the quadrature points are valuable outputs of the workflow to enable humans to assess the benchmarking results.

\section{A general framework}
\label{sec:general-framework}
\deleted{\Teoman{I would swap the position of this section with the previous one.}\\}
In the previous \replaced[id=TT]{s}{S}ection\comment[id=ML]{We should unify this. Section is now sometimes written with lower and sometimes with upper cases.}, we have described our very specific use case.
In this section, we are going to derive a general framework from this use case,
which the reader might apply to their own case.

In an abstract view, 
one faces a set of test cases $T = \{ \tau_1, \ldots, \tau_n \}$
and a set of algorithms $A = \{ a_1, \ldots, a_m \}$.
The goal is to evaluate the algorithms on the test cases and to draw some measures,
which one might denote as a mapping $m(\tau, a)$, 
which assigns a measurement result, e.g., a real number,
 to each pair of test case and algorithm $(\tau_i, a_j)$ in the Cartesian product $T \times A$.

 \paragraph{An abstraction layer for all libraries to test}
For the organization of the benchmarking, 
it is crucial to define a common software interface,
test cases $T$, algorithms $A$, and the mapping $m$.

If an object-oriented approach, resp. programming language, is employed, 
this common software interface can be defined as abstract classes; 
some programming languages also directly support interfaces.
In Python for example, one could use abstract base classes for this using the \texttt{@abstractmethod} decorator,
which enforces that all derived classes implement the same methods.
In MATLAB, one could use a similar approach by defining an abstract class (\texttt{classdef (Abstract)}) with abstract methods. 

Alternatively, in certain languages like Python, it is not necessary to define an abstract class,
but one could also use so-called \emph{duck typing}:
This means that the common interface is not enforced by the language, but is rather a convention,
i.e., all classes that are supposed to be used in the benchmarking 
must have the same methods with the same signatures.
If a class does not have the required methods, Python will raise an exception at runtime.

If a procedural approach is used instead of an object-oriented approach,
a common software interface can be defined by a set of functions,
very often called \emph{driver routines},
which all have the same signature.

\paragraph{The choice of scripting language}
As already mentioned, we have chosen MATLAB for our use case.
Obviously, there are many alternatives.
A safe alternative choice, in the domain of scientific computing, would certainly be Python\added[id=TT]{, also considering licensing}.
Depending on the specific case, also Julia, Ruby or R might be considered.

Nevertheless, the following criteria should be considered when choosing a scripting language:
At first, one should prefer some interpreted language that works on all envisioned platforms, resp. operating systems.
We should note here that the important criterion is that the code should work on all platforms without 
re-compilation, which is required, e.g., for C code.
Languages like MATLAB or Julia \replaced[id=TT]{also}{certainly} fulfill this criterion,
although they are not interpreted languages in the strict sense.
Technically, they are so-called Just-In-Time (JIT) compiled languages, which are compiled at startup or even in part at runtime, 
but this is rather invisible to the user.

A very important criterion for the choice of the scripting language 
might be the support for some level of object orientation,
which is a great aid in creating common interfaces.

In scientific computing,
a further, often overlooked criterion is the support for creating visualizations.
This capability is especially important in the context of benchmarking where plots are helping
to assess results. Additionally, the plotting solution of choice should also 
enable the creation of high-quality vector-graphics plots
which can be directly included in (\LaTeX) documents. For example, Python with \texttt{matplotlib}\footnote{\url{https://matplotlib.org/}, accessed: 21.03.2025} and MATLAB\footnote{\url{https://uk.mathworks.com/help/matlab/graphics.html}, accessed: 21.03.2025} for example
provide powerful plotting tools where vector-output can be created using the \texttt{pgf} backend or the 
\texttt{print} command, respectively.


Obviously, none of the criteria discussed above are a must-have, but the lack of some of them would rule out a language.
A general recommendation is, however, to keep the benchmarking scripts as minimal and simple as possible.
Therefore, it is wise to choose some language that provides all or most required features out-of-the-box,
without the need for cumbersome workarounds.
\added[id=TT, comment={I guess this suffices?}]{Moreover, storing the underlying data in a neutral format helps to create or manipulate future visualizations.}
\comment[id=ML]{A further feature, which we used and which is helpful, is the generation, storing and handling of tables. It's minor than the already mentioned points. Shall we add that?}

\paragraph{Using submodules}
We strongly advocate using \emph{Git submodules} for including third-party libraries. 
Submodules are a feature of Git, allowing a repository to be a subdirectory of another repository
(\emph{main repository}).
Since most software projects use nowadays Git for version control,
this can be done for almost any third-party library.

Using submodules gives the following advantages: First, from a legal point of view, one is free to publish \emph{his} \emph{main repository} under any terms
without being bound to the licenses of the submodules.
If someone clones the \emph{main repository},
this other person does not obtain their copy of the third-party libraries/submodules from the author of the \emph{main repository},
but from the original authors.
Second, one can track which version of the third-party libraries is used
since the commits of the main repository are linked with specific commits of the submodules.
Third, one can easily update the submodules to newer versions of the third-party libraries, if one wishes to do so.
Fourth, the \emph{main repository} remains cleaner and lighter. This is especially advantageous if multiple large libraries should be linked, e.g., as in our use case.

\comment[id=ML]{Even though I also prefer submodules, I also see possible disadvantages compared to hard copies which we could list here. 
Dependency: If a repository gets deleted, the code is not available even though it was open-source before (similar if they move the repo or change the address). 
User friendliness: We also had some "trouble" to link all submodules correctly; with a hardcopy, the user only has to take care of cloning the "main repository".}
\paragraph{Alternative to submodules -- using subtrees}\comment[id=TT]{Does following section address your concerns?}
Obviously, submodules also come with some disadvantages.
One major concern is dependency: If a repository gets deleted, the code is not available even though it was open-source before.
Even if the repository is just moved, the initialization of submodules
will not work anymore since the link is broken.
Furthermore, submodules also require additional Git commands to be initialized and updated correctly.
It might therefore be more practical to directly include the third-party libraries into the main repository,
given that this is not prohibited by the license of the third-party libraries.

When the files are just inserted by a copy operation, the entire Git history of the third party is lost. In that case, the individual who copied the respective files is shown as the original author in the git history.
This is unaesthetic, at least with regard to the recognition of authorship, 
although this may be mentioned separately in the files.
However, it is also not practical if the version history of a file is lost.

In order to avoid this, one can use \emph{Git subtrees} instead of plain copying of files.
A subtree is a subdirectory of a repository that has its own history. In other words, subtrees allow merging two distinct repositories into one,
while maintaining the history of both.

\paragraph{Setting up a CI workflow}
A CI workflow is a structured process to integrate changes into the shared repository. In general, there are five stages: code integration, automated builds, automated tests, analysis, and review (merge request), supported by version control to track changes. 

Each contributor performs their\comment[id=ML]{If we use the gender neutral form here which I like. We should do it everywhere which is not the case so far.} changes in their local branch and resp. repositories until it is mature enough to merge with the \emph{main branch}. 
\replaced[id=ML]{The m}{M}ain branch is designated to serve as the working and latest state of the codebase and is protected from direct changes without successful pipelines. 
After a merge request is created, the platform automatically tries to build the software with newly introduced changes in pre-defined environments (e.g., remote servers with different operating systems) through runners. 
Hence, it initially checks for fundamental problems such as missing dependencies and compiling errors. 
This particular step ensures that basic requirements for the software are provided, such as third-party libraries or correct syntax. 

If all the build tests succeed, 
the pipeline proceeds with the execution of automated tests on these built environments. 
These tests are designed to verify individual components (i.e., unit testing), 
and validate the overall behavior of the software at the current state (i.e., regression testing).
Moreover, the tests should provide sufficient coverage across heterogeneous architectures (e.g., operating systems or versions) to ensure the software behaves consistently and reliably in various scenarios.

In this stage, we propose that each algorithm in $A$ initially undergoes quick tests with relatively faster results. 
It is observed that an early detection of such problems has a positive impact on developers, 
leading to a more agile process and a reduction in the computational cost associated with more extensive tests in case of failures. 
It should be noted that these quick cases, consisting of a minority of the cases defined in $T$, focus not only the working status, 
but also check the expected values under predefined tolerances.\comment[id=ML]{The second part of the foregoing sentence is not clear to me. Maybe, you can reformulate it or even delete it.}
After each component is verified in the fast test environment, 
more extensive tests are conducted, e.g., numerical convergence tests, also with the possibility of combining multiple or all components if desired. 
Reproducing published or established values as test cases also aids to determine potential bugs introduced with changes and ensures backward comparability. 

Upon successful completion of the pipeline, a merge request receives the successful status and is listed as ready. 
Then, the contributors or supervisors have the opportunity to analyze the changes, allowing for manual quality assurance. 
This review process becomes particularly important with larger iterations or changes considering components maintained by others.
When the contributors approve the changes, the request is merged into the main branch, updating the remote repository. 
\deleted[id=ML]{It should be noted that the approval review can also be skipped or accepted in advance in the case of GitLab pipelines.}

Another important detail in the CI workflow is the automatic collection of the artifacts generated during tests. 
With artifacts, not only the status of test cases but also the actual results become visible and trackable. 
A developer can, therefore, analyze the actual behavior of the proposed changes. 
Moreover, these artifacts should be uploaded and stored in the remote repository.
In our experience, a two-days period suffices the need for this purpose, with an option to download or keep the artifact \replaced[id=ML]{stored longer on the remote server}{on the remote server stored longer}.
Additionally, we propose to produce artifacts in neutral and human-readable formats. 
A neutral format enhances the research data exchange and management for later usages, while having also a browser supported summary (e.g., HTML or PDF) significantly promotes the understanding of developers.
This human-readable summary can also be formulated like automated reports with minor efforts.  

Beyond unit and regression testing, a possible type of testing is performance testing. Depending on the benchmark and project goals, performance testing can play a vital role by enabling the monitoring of key metrics like resource consumption and scalability. Therefore, it provides key insights into the software behavior on different computational architectures and helps the identification of bottlenecks. 

Furthermore, we advocate using local testing environments or pipelines independent of merge requests. 
It is observed that having an option to privately test changes prior to a merge request reduces developers' anxiety over a potential failure, especially in heterogeneous and unfamiliar research environments. 
In addition, it helps to determine conflicts and underlying issues in case of concurrent changes by multiple developers. 

Moreover, establishing a local testing environment ensures consistency between developers' setups and the CI pipeline, minimizing discrepancies during integration.
Testing within CI pipelines is operated by independent agents (e.g., GitLab runners).
These runners actively listen to the remote server, downloads the codebase with changes and execute tests, when a pipeline is triggered.  
To efficiently achieve this, runners should be configured to closely replicate the target testing conditions and possess the necessary resources and diversity (e.g., CPU count, operating system types).

\paragraph{The usage of containers}
Within recent years, the use of \emph{container} solutions, such as Docker, 
has become increasingly popular.
In a nutshell, 
a container is a lightweight, standalone, executable package that includes 
everything needed to run an application: 
operating system, 
required system libraries and tools, the respective application of interests and its configuration  settings.
This can drastically simplify the distribution of software:
For any software, it is usually very difficult to ensure that it runs on all possible operating systems.
Even on Linux, it is difficult to create a binary package which runs on all major distributions out-of-the-box.
With containers, however, this is much easier:
One can create a container which contains the required environment and the software,
and this container can be executed on any host system which supports the container technology.

One might also think about containers as lightweight virtual machines.
Virtual Machines (VMs) are independent installations of an operating system
which can be executed on a host system using a software such as VirtualBox or VMware.
A classical virtual machine, however, must be maintained as any other operating system installation,
e.g., it must be updated regularly.
It also takes a significant time to boot, like any other operating system.
Furthermore, while virtual machines can be transferred from one host system to another,
this is not always straightforward. It is also untypical, e.g., to put a virtual machine image up for download on a website.

Containers, on the other hand, are much more lightweight.
The operating system is as minimal as possible and time to start a container is negligible.
Furthermore, systems like Docker are explicitly designed to be used in a cloud environment,
i.e., containers can be \emph{pushed} to online repositories and \emph{pulled} from there
by any other user.

With respect to GitLab runners or other CI systems,
it can be helpful to use containers to run test jobs.
This ensures that the tests are always executed in the same, well-defined environment.
E.g., if some team decides to use its workstations also for executing tests,
one can change the installation of any workstation, without the risk of breaking the tests.

Especially for Docker, this goes even further:
It is not even necessary to distribute container images, 
which might have a size of several megabytes up to gigabytes,
depending on the software environment in the container.
Instead, one can use \emph{Dockerfiles},
which are small scripts to build a container image from scratch.
These Dockerfiles can be conveniently stored and maintained in a Git repository.

The use of containers, however,
does not come without some potential pitfalls:
In containers, it is easy to install software that is not maintained anymore, 
e.g., old versions of libraries.
Then, especially small development teams might fall for the temptation of operating with `time-capsules',
i.e., of not updating their code to work with recent versions of libraries.
This can lead to a situation where the code is, 
at some point, impossible to run on recent operations systems.
On the other hand, such time-capsules
at least allow the reproduction of results, 
several years after the original publication,
when the research code was not further maintained and the original team does not exist anymore.

Finally, one should also consider the licensing of the software in the container:
Due to the anonymous online distribution of containers,
this prohibits the use of non-open-source software in containers,
e.g., MATLAB.

\comment[id=TT]{We did not use containerazation but for the sake of arguments here, one can introduce them. The problem is to ensure licensing with MATLAB in Docker.}



\section{Conclusion}
\label{sec:conclusion}

In this work, Continious Integration (CI) is successfully applied to benchmarking research software with the use case on numerical integration in cut elements. 
By employing this technique, 
it is aimed to facilitate scientific benchmarking in accordance with state-of-the-art tools. 
From this experience, a comprehensive set of principles and practices applicable to more general settings is derived and presented.

In our specific use case with the numerical integration, 
it is observed that using a platform employing CI techniques streamlines the comparative analyses with established references and previous versions. 
Hence, it not only enables systematic monitoring of changes or iterations,
but also boosts software quality and development efforts. 
Furthermore, it ensures a reliable and stable version of the benchmark associated with scientific methods and promotes compatibility of results across diverse computational and research environments.

In our framework, we propose a unified driver interface for components (e.g., libraries, unit features) and tests, serving as an abstract layer. 
As this layer plays a central role in integrating components, 
we recommend adopting an interpreted language that supports an object-oriented paradigm and visualization capabilities to facilitate integration and analysis.
Additionally, designing a testing suite that initially executes rapid tests is observed to increase overall efficiency by quickly identifying potential issues before proceeding to more extensive and computationally intensive validations.

Along with automated pipelines, 
it is also observed that using version controlling and containers offers valuable benefits.
Version controlling systems, such as Git, provide useful features for both developers and testing platforms. 
For example, employing third-party libraries can be seamlessly managed with submodules and can prevent potential licensing issues.
Moreover, consistent and reliable testing environments can be achieved through containers, ensuring the same  requirements for computational infrastructure.

In addition, CI-driven benchmarking inherently promotes a better scientific practice, adhering to the FAIR principles. 
Consequently, 
scientific research becomes more transparent, reliable, and reproducible, creating a lasting digital record and accountability in research.
In the future, researchers can utilize the proposed framework using similar tools to establish benchmarks for common problems. 
This would not only aid them to develop their own work but also serve as a common basis for researchers across different domains, ensuring consistency and comparability over time.







\section*{Acknowledgment}
The work of T. Toprak, G. Teixeira, I. Shiskina, F. Kummer and B. Marussig is partially supported by the
joint DFG/FWF Collaborative Research Centre CREATOR (DFG: Project-ID
492661287/TRR 361; FWF: 10.55776/ \allowbreak F90) at TU Darmstadt, TU Graz and
JKU Linz. Additionally, the work of T. Toprak, I. Shiskina, C. Miao and
F. Kummer is supported by the Graduate School CE within the Centre for
Computational Engineering at TU Darmstadt.

\section*{Data availability}
The data is available from \cite{CutElementIntegration2025}.
\section*{Declaration of competing interest}
The authors declare that they have no known competing financial interests or personal relationships that could have appeared to influence the work reported in this paper.
\section*{CRediT authorship contribution statement}
\textbf{Teoman Toprak}: Conceptualization, Investigation, Methodology, Software, Visualization, Writing – original draft, Writing – review \& editing. \textbf{Michael Loibl}: Conceptualization, Investigation, Methodology, Software, Visualization, Writing – original draft, Writing – review \& editing.
\textbf{Guilherme H. Teixeira}: Conceptualization, Investigation, Methodology, Software, Visualization, Writing – original draft, Writing – review \& editing. \textbf{Irina Shishkina}: Investigation, Writing – review \& editing.
\textbf{Chen Miao}: Software, Visualization, Writing – review \& editing.
\textbf{Josef Kiendl}: Writing – review \& editing. 
\textbf{Benjamin Marussig}: Conceptualization, Funding acquisition,  Methodology, Resources, Software, Supervision, Writing – review \& editing.  \textbf{Florian Kummer}: Conceptualization, Funding acquisition, Project administration, Writing – original draft, Writing – review \& editing.




  \bibliographystyle{elsarticle-num} 
  \bibliography{mybibliography}






\end{document}